\documentclass[aps,prd,twocolumn,superscriptaddress,nofootinbib,floatfix]{revtex4-2}

\usepackage{amsmath,amssymb,bm}
\usepackage{graphicx}
\usepackage{array,dcolumn,multirow}
\usepackage{xcolor}
\usepackage{comment}
\usepackage{orcidlink}
\definecolor{myred}{RGB}{179, 27, 27}
\usepackage{hyperref}
\hypersetup{
    colorlinks=true,
    linkcolor=myred,
    citecolor=myred,
    urlcolor=myred
 }

\newcommand{\MBH}{M_{\rm BH}}
\newcommand{\Comp}{\mathcal{C}}
\newcommand{\Zh}{Z_h}
\renewcommand{\Re}{\operatorname{Re}}
\renewcommand{\Im}{\operatorname{Im}}

\begin{document}

\title{Quasinormal modes of black holes in astrophysical environments:\\ Rainbow in the dark matter}

\author{Kyriakos Destounis \orcidlink{0000-0003-2997-088X}}
\affiliation{CENTRA, Departamento de F\'isica, Instituto Superior T\'ecnico -- IST, Universidade de Lisboa -- UL, Avenida Rovisco Pais 1, 1049-001 Lisboa, Portugal}
\affiliation{School of Applied Mathematical and Physical Sciences -- SEMFE, Physics Department, National Technical University of Athens -- Polytechnic Campus, Zografou 15772, Athens, Greece}
\author{Laura Pezzella \orcidlink{0009-0008-6357-6417}}
\affiliation{Gran Sasso Science Institute (GSSI), I-67100 L'Aquila, Italy}
\affiliation{INFN, Laboratori Nazionali del Gran Sasso, I-67100 Assergi, Italy}
\begin{abstract}
Matter around black holes affects the ringdown in two ways: through the background geometry and through the response of the matter itself. We compute axial and polar gravitational quasinormal modes of a black hole embedded in a relativistic Hernquist-type dark matter halo, following the fundamental mode and the first three overtones associated with the vacuum Schwarzschild spectrum. The polar sector couples metric and matter perturbations. At small halo compactness, both sectors follow the familiar gravitational redshift law. Most of the departure from the typical redshift at larger compactness is accounted for by the exact horizon lapse factor of the background. The remaining residual departure to this lapse is sector dependent. Axial and polar frequencies are split, with their splitting growing from the fundamental through the third overtone. At large halo compactness the split grows by an order of magnitude, while overtone-to-fundamental frequency ratios depart from a common rescaling. Within the adopted dark matter profile, the halo acts on the ringdown as a prism that disperses the gravitational spectrum, leaving a parity-dependent imprint that no common rescaling of the frequencies can reproduce. Parity-isospectrality is, thus, broken.
\end{abstract}

\maketitle

\section{Introduction}

Much as optical spectroscopy identifies a source from the colors of its light, black-hole (BH) spectroscopy \cite{Berti:2025hly,Destounis:2023ruj} reads the geometry of a perturbed remnant off its damped oscillations, namely the quasinormal modes (QNMs) \cite{Dreyer:2003bv,Kokkotas:1999bd,Berti:2009kk,Konoplya:2011qq}. The high signal-to-noise BH merger GW250114~\cite{LIGOScientific:2025rid,LIGOScientific:2025wao} illustrates the prospects for testing general relativity with the ringdown~\cite{LIGOScientific:2016lio,Berti:2015itd,Franchini:2023eda}. This analysis found evidence for the fundamental QNM and the first overtone under the models considered in \cite{LIGOScientific:2025wao}. Related studies have explored nonlinear QNMs \cite{Cheung:2022rbm,Mitman:2022qdl,Giesler:2024hcr,Ma:2024qcv}, direct waves \cite{Ma:2023cwe,Lu:2025vol,Dyer:2026sox,Dyer:2026yex,Ma:2026hcb,Cheung:2026gfd,Nakamoto:2026rah}, exceptional points \cite{Heiss:2012dx,Cavalcante:2024swt,PanossoMacedo:2025xnf,Yang:2025dbn,Cao:2025afs,Zhang:2026uoo,Rossi:2026als,Destounis:2026qne,Destounis:2026xvd}, and spectral instabilities through mathematical tools of non-modal analysis \cite{Nollert:1996rf,Jaramillo:2020tuu,Jaramillo:2021tmt,Jaramillo:2022kuv,Cheung:2021bol,Courty:2023rxk,Destounis:2025dck,Cardoso:2024mrw,Berti:2022xfj,Yang:2024vor,May:2026qpf,Destounis:2021lum,Boyanov:2022ark,Boyanov:2023qqf,Destounis:2023nmb,Arean:2023ejh,Cownden:2023dam,Cai:2025irl,Zhang:2026cjq,Destounis:2026ays}.

Astrophysical BHs can be surrounded by matter that alters how gravitational waves are generated and propagate. This raises the question of how does the astrophysical environment that surrounds BHs affect the ringdown signal and its QNM content. The relativistic BH--halo solution of Ref.~\cite{Cardoso:2021wlq,Cardoso:2022whc}, together with the analysis in Ref. \cite{Pezzella:2024tkf}, has shown that axial QNMs are collectively changed by a common factor. Extensions of the solution in \cite{Cardoso:2021wlq} to other halo profiles \cite{Konoplya:2022hbl,Jusufi:2022jxu}, and central BH geometries \cite{Feng:2022evy,Stelea:2023yqo,Fernandes:2025osu,Yue:2026gju} have been recently found. These solutions have also been used in extreme-mass-ratio setups, where the fluxes and halo detectability has been studied  \cite{Cardoso:2022whc,Destounis:2022obl,Speeney:2024mas,Destounis:2025tjn,Datta:2025ruh,Gliorio:2025cbh,Zhao:2026eti}. For a sparse halo, the dominant effect on the axial QNMs is a gravitational redshift \cite{Pezzella:2024tkf,Berti:2025hly}. In the polar sector, though, metric perturbations couple with matter, and the fluid motion feeds back on the metric. 

A rainbow offers a useful picture of the difference between these two effects. In vacuum, all colors of sunlight propagate together, and a medium with the same refractive index at every wavelength would delay and bend them uniformly, leaving white light white. A rainbow forms only because water droplets refract different wavelengths by different amounts, i.e., because the medium is dispersive. The Schwarzschild spectrum plays the role of white light: its axial and polar sectors are isospectral \cite{Chandrasekhar:1975zza,Chandrasekhar:1985kt}. A gravitational redshift is the analog of a non-dispersive medium, since it rescales all modes by a common factor and preserves isospectrality. The matter response in the polar sector, which the axial sector lacks, acts instead like the droplets, giving the two parities, and successive overtones, different responses to the same halo. Dark matter does not disperse light, but the question we address is whether it disperses the gravitational ringdown, producing a \emph{dark matter rainbow}\footnote{The analogy used regards only spectral dispersion, not propagation}.

We adopt a relativistic Hernquist-type halo with constant radial sound speed, vanishing tangential sound speed and no inner matter cutoff. The axial and polar fundamental-mode parameter space search covers the halo masses $M/\MBH=10,10^3$, where $\MBH$ is the central BH mass, $\ell=2,3,4$ and compactness $10^{-6}\le\Comp\le10^{-1}$, where $\Comp=M/a_0$, with $a_0$ the halo's length scale. The scan is supplemented by $\ell=2$ runs at smaller halo mass and by a scan in sound speed. We then compare axial and polar modes for the first three overtones, $n=1,2,3$, at $M/\MBH=1,10$, $c_{s,r}=0.9$ and five compactness values between $10^{-3}$ and $10^{-1}$. We find that the common gravitational redshift of axial QNMs approximately survives in the polar sector when the compactness is very small, though parity-dependent imprints appear for more compact dark matter halos that no typical rescaling can replicate. Therefore, parity-isospectrality is broken. Throughout this work we use geometrized units $G=c=1$.

\section{Background and perturbation equations}\label{sec:background}

\subsection{Relativistic halo geometry}

We use the Hernquist-type BH geometry constructed in Ref.~\cite{Cardoso:2021wlq},
\begin{equation}
ds^2=-a(r)\,dt^2+\frac{dr^2}{b(r)}+r^2d\Omega^2_2,
\end{equation}
with
\begin{align}
a(r)&=\left(1-\frac{r_h}{r}\right)e^{\Upsilon(r)},\qquad r_h=2\MBH,\\
\Upsilon(r)&=-2\sqrt{\frac{M}{\xi}}\arctan\left(\frac{\sqrt{M\xi}}{r+a_0-M}\right),\label{eq:Upsilon}\\
\xi&=2a_0-M+4\MBH.
\end{align}
The radial metric function and the enclosed mass are \cite{Cardoso:2021wlq}
\begin{equation}
b(r)=1-\frac{2m(r)}{r},\quad m(r)=\MBH+M\frac{(r-r_h)^2}{(r+a_0)^2},
\end{equation}
respectively, while the energy density of Hernquist-type halos follows from
\begin{equation}
\rho(r)=\frac{m'(r)}{4\pi r^2},\qquad m'(r)=\frac{2M(r-r_h)(a_0+r_h)}{(r+a_0)^3}.
\end{equation}
Here, primes denote $d/dr$ throughout. The Arnowitt-Deser-Moser (ADM) mass is $M_{\rm ADM}=\MBH+M$, and the vacuum limit is Schwarzschild. At finite $\MBH$ the density vanishes at the horizon, i.e., as $r_h\to0$ it approaches the Newtonian Hernquist profile \cite{Hernquist:1990be}. The equilibrium anisotropic stress-energy tensor is $T^\mu{}_\nu={\rm diag}(-\rho,0,p_t,p_t)$, as motivated by the Einstein cluster description \cite{Einstein:1939ms}, with $p_t=m\rho/[2(r-2m)]$. We set $f=\sqrt{ab}$ so that
$d/dr_*=f\,d/dr$. The identities
\begin{equation}
\frac{a'}{a}=\frac{1-b}{rb},\qquad
D\equiv rb'+b-1=-2m',
\label{eq:identities}
\end{equation}
enter the perturbation equations below. The horizon lapse
\begin{equation}
\Zh=e^{\Upsilon(r_h)/2}
=1-\Comp+\left(\frac16+\frac{r_h}{M}\right)\Comp^2+O(\Comp^3)
\label{eq:Zh}
\end{equation}
sets a redshift benchmark for QNMs by rescaling the near-horizon timescale. At fixed $M$ and $r=O(\MBH)$, local matter terms start at order $\Comp^2/M$. This hierarchy provides a useful context for the nearly quadratic splitting at low compactness, but it is not an exact QNM scaling law.

Extending the stress tensor to the horizon is phenomenological: $p_t$ remains finite and nonzero there even though $\rho$ vanishes, and the timelike circular-orbit interpretation requires $r>3m$. The more compact examples test the model and are not meant to represent typical galactic dark matter halos.

\subsection{Axial sector}

We decompose the metric perturbation in spherical harmonics such that $\Psi(r) Y_{\ell m}(\theta,\phi)e^{-i \omega t}$, where $e^{-i \omega t}$ is our choice for the harmonic time dependence throughout this work. The homogeneous, radial, axial master equation becomes \cite{Cardoso:2022whc}
\begin{align}\label{eq:axial_equation}
&\frac{d^2\Psi}{dr_*^2}+\left[\omega^2-V_{\rm ax}(r)\right]\Psi=0,\\
&V_{\rm ax}(r)=\frac{a}{r^2}\left[\ell(\ell+1)-\frac{6m}{r}+m'\right].
\end{align}
In the radial coordinate $r$ Eq.~\eqref{eq:axial_equation} becomes
\begin{align}
&\Psi''+P\Psi'+Q\Psi=0,\\
&P=\frac{f'}{f}=\frac12\left[\frac{1-b}{rb}+\frac{b'}{b}\right],\qquad Q=\frac{\omega^2-V_{\rm ax}}{ab}.
\end{align}
The potential has the form associated with the covariant-velocity prescription of Ref.~\cite{Chakraborty:2024gcr}, which distinguishes fixed covariant from fixed contravariant fluid perturbations on anisotropic backgrounds. We use this potential for the baseline comparisons with Ref.~\cite{Pezzella:2024tkf}. Appendix \ref{app:common_tilt} gives the shared angular condition for the baseline axial and polar operators.

\subsection{Polar sector}

Polar perturbations couple the metric sector to matter~\cite{Cardoso:2022whc}. We write the anisotropic stress tensor as
\begin{equation}
T_{\mu\nu}=(\rho+p_t)u_\mu u_\nu+p_tg_{\mu\nu}+(p_r-p_t)k_\mu k_\nu,
\end{equation}
with $u^\mu u_\mu=-1$, $k^\mu k_\mu=1$ and $u^\mu k_\mu=0$.
The Eulerian pressure closure is
\begin{equation}
\delta p_r=c_s^2\delta\rho,\qquad \delta p_t=0,
\label{eq:closure}
\end{equation}
with constant $c_s=c_{s,r}$ and $c_{s,t}=0$. These relations prescribe the pressure response independently of the equilibrium
profiles. The choice $c_{s,t}=0$ retains the nonzero background tangential pressure. Separately, in the baseline Regge--Wheeler
formulation we set the independent angular tilt $\delta k_A$ to zero, where $A\in\{\theta,\phi\}$. The anisotropy vector remains nonzero, while its time and radial perturbations are fixed by normalization and orthogonality.

We write $\lambda=-i\omega$ and $\bm{X}=(H_1,H_0,K,W,\delta\rho)^T$. The first three components are metric amplitudes in our Regge--Wheeler-gauge convention and $W$ is a radial fluid variable. Its normalization relative to the perturbed four-velocity, the metric-harmonic and the angular-fluid conventions are given in Appendix \ref{app:physical_audit}.

The polar frequencies are found by solving the following system (after applying appropriate boundary conditions)
\begin{equation}
\bm{X}'=F(r,\lambda)\bm{X},\qquad \bm{c}(r,\lambda)\bm{X}=0,\label{eq:polarsystem}
\end{equation}
whose nonzero entries of $F$ are listed in Appendix~\ref{app:matrix}. With $L=\ell(\ell+1)$, $s=c_s^2$ and $D=-2m'$, the constraint row is
\begin{align}
c_{H_1}&=\frac{La(b-1)}{2\lambda r}-2\lambda rb,\label{eq:cH1}\\
c_{H_0}&=a(3b-1-L),\\
c_K&=a(L-2)+2\lambda^2r^2-\frac{a(b-1)(3b-1)}{2b},\\
c_W&=\frac{8ra^2(b-1)}{\lambda b(3b+1)},\\
c_{\delta\rho}&=16\pi r^2as.\label{eq:cdrho}
\end{align}
Near the BH we eliminate $H_0$ and evolve $\bm{z}=(H_1,K,W,\delta\rho)^T$ after the rescaling
\begin{equation}
\bm{y}=T\bm{z},\qquad T={\rm diag}(r^{-1},1,r^2,r^2).
\end{equation}
Writing $\mathcal{I}=(H_1,K,W,\delta\rho)$, the reduced matrix is
\begin{equation}
\bar F_{\mathcal{IJ}}=F_{\mathcal{IJ}}-F_{\mathcal{I}H_0}\frac{c_\mathcal{J}}{c_{H_0}},\qquad B=T\bar FT^{-1}+T'T^{-1}.\label{eq:reduced}
\end{equation}
Reconstructing $H_0$ from the constraint enforces it algebraically, and Appendix \ref{app:constraint} shows that the reduced evolution also satisfies the omitted differential equation of the five-component system.

\section{Boundary conditions}\label{sec:bcs}

In our model, we extend the background and the constant radial sound speed over $r_h<r<\infty$. We choose retarded solutions at the horizon and impose no incoming radiation from infinity. The polar acoustic condition is independent of the gravitational one.

\subsection{Axial sector}

We set $x=r-r_h$ and $\tau_h=\sqrt{r_h/a'(r_h)}=r_h/\Zh$.
Up to additive constants, $r_*\sim\tau_h\ln x$ at the horizon and $r_*=r+2M_{\rm ADM}\ln r+O(r^{-1})$ at infinity. With time dependence $e^{-i\omega t}$, the ingoing and outgoing axial series are, respectively,
\begin{align}
\Psi_{\rm in}
 &= x^{-i\omega\tau_h}\sum_{n=0}^{N_h}a_n x^n,\\
\Psi_{\rm out}
 &= e^{i\omega r}r^{2iM_{\rm ADM}\omega}
    \sum_{n=0}^{N_\infty}\frac{b_n}{r^n}.
\end{align}
Here, $N_h$ and $N_\infty$ are the truncation orders. We obtain the fields and derivatives from these series.

\subsection{Polar horizon: gravitational and acoustic branches}

In the rescaled fields $\bm Y_h=(xH_1,xH_0,K,W,\delta\rho)^T$, the near-horizon radial matrix has the expansion $\sum_{n=0}^{\infty}R_n x^{n-1}$, where $R_n$ is the corresponding coefficient matrix. Substituting a leading solution $x^{p_{j}}\bm v_0$ gives the indicial equation $R_0\bm v_0=p_j\bm v_0$. Here, $j=\{g,f\}$ serves as an index that provides a distinction between the gravitational exponent $p_g$ and the fluid component $p_f$. The two retarded exponents are
\begin{align}
p_g&=-i\omega\tau_h=\lambda\tau_h,\\\label{eq:pf}
p_f&=-\frac{3s+1}{4s}+\nu, \quad
\nu^2=\left(\frac{1-s}{4s}\right)^2+\frac{\tau_h^2\lambda^2}{s}.
\end{align}
The acoustic branch is defined by $\nu>0$ for real $\lambda>0$ and then by analytic continuation. In the domain used here this gives
\begin{align}
\nu&=-i\sqrt{(\tau_h\omega/c_s)^2-\delta^2},
\label{eq:nu}\\
\delta&=\frac{1-s}{4s},\qquad
\omega_{\rm th}=\frac{1-s}{4c_s\tau_h}.
\end{align}
We take the principal square root, approaching the positive imaginary $\omega$ axis from its right and crossing the real axis at $\omega>\omega_{\rm th}$, where the square-root argument is positive. Our searches use $\Re\omega>1.05\,\omega_{\rm th}$, where this margin avoids the branch point and does not exclude physical roots elsewhere.

The starting column is $\bm Y_h^{(j)}=x^{p_j}\sum_{n=0}^{N_h}\bm v_n^{(j)}x^n$,
with coefficients determined by
\begin{equation}
\big[(p_j+n)I-R_0\big]\bm v_n^{(j)}
=\sum_{k=1}^{n}R_k\bm v_{n-k}^{(j)} .
\end{equation}
Appendix~\ref{app:horizon} provides the leading vectors. The inner solution is an arbitrary combination of the two coupled
columns. The acoustic branch specified here is a formal radiation condition: the matched endpoint modes have a nonzero acoustic component whose density diverges in ingoing coordinates (see Appendix \ref{app:acoustic_audit}).

\subsection{Polar infinity: outgoing waves and sound}

The metric and acoustic waves decouple at leading order at infinity. Their outgoing phases are $e^{i\omega r_*}$ and $e^{i\omega r_*/c_s}$, respectively. In the rescaled variables $\bm u =(H_1/r,H_0/r,K,r^2W,r^2\delta\rho)^T$, the outgoing solutions are
\begin{align}
\bm u_j
 &=e^{k_j r}r^{\eta_j}
   \sum_{n=0}^{N_\infty}\frac{\bm v_{jn}}{r^n},
 & \eta_j&=2M_{\rm ADM}k_j,\\
k_g&=i\omega, & k_f&=i\omega/c_s .
\end{align}
The factor $r^{\eta_j}$ follows from the logarithmic term in $r_*$. The leading field equations give, up to overall normalization,
\begin{align}
\bm v_{g0}&=(\lambda,-\lambda,1,0,0)^T,\\
\bm v_{f0}&=(0,0,0,4\pi c_s,1)^T .
\end{align}
Higher orders mix the metric and matter fields. The outer condition keeps both outgoing columns and discards both incoming ones. At the finite starting point we evaluate the two outgoing series without imposing $\delta\rho=0$ there.

\section{Numerical method}
\label{sec:numerics}

\subsection{Boundary expansions and shooting method}

We find the QNMs by shooting from both boundaries and matching at a common real radius $r=r_m$. At a trial frequency, the axial problem contributes one solution from each side, while the polar problem contributes two, one gravitational and one acoustic. Their relative amplitudes are free until the match.

We generate the starting series with 45 decimal digits for the axial expansions and 50 for the polar expansions, then use complex double precision for integration and matching. The polar infinity recurrence uses one extra order: at order $n$ the leading radial matrix minus $k_j I$ is singular, so its solvability condition fixes the previously undetermined component of $\bm v_{j,n-1}$ along $\bm v_{j0}$. The first such condition checks $\eta_j=2M_{\rm ADM}k_j$. We reject ill-conditioned boundary systems and resonant horizon recurrences.

The inner solutions are integrated along the real exterior with an adaptive eighth-order Runge--Kutta method. We periodically QR-orthonormalize the polar columns to prevent
one from masking the other while preserving their span. We integrate the outer solutions inward along $r=r_m+e^{i\theta}t$, $0\le t\le L_{\rm out}$, choosing $\theta$ so that the outgoing exponentials decay toward the
outer endpoint. For long contours we use implicit backward differentiation in the far region. The axial variable is the logarithmic derivative $q=\Psi'/\Psi$, satisfying $q'=-q^2-Pq-Q$. In the far region of the polar calculation, we use the constraint to eliminate $K$ rather than $H_0$. We rescale the four remaining fields as $\bm u=(H_1/r,H_0/r,r^2W,r^2\delta\rho)^T$. Their radial equation is $\bm u'=G_\infty\bm u$, where $G_\infty$ includes the derivatives of the radial rescaling factors. Appendix \ref{app:polar_riccati} gives the explicit construction.

At infinity, the leading equations separate into four waves: outgoing gravitational, outgoing acoustic, incoming gravitational, and incoming acoustic. We use their leading vectors, in that order, as the columns of the matrix $S_\infty$ given in Appendix \ref{app:polar_riccati}. Writing $\bm u=S_\infty\bm w$ expresses the same four fields in these wave coordinates. At a fixed trial frequency $S_\infty$ is independent of $r$, so the radial equation becomes
$$
\bm w'=S_\infty^{-1}G_\infty S_\infty\bm w.
$$
Splitting the coordinates after the first two gives
\begin{equation}
\mathcal M\equiv S_\infty^{-1}G_\infty S_\infty
=\begin{pmatrix}
 A & B_{12}\\
 C & D_4
\end{pmatrix}.
\end{equation}
Each block is $2\times2$. The diagonal blocks act within the outgoing and incoming coordinates, respectively, and the off-diagonal blocks couple them at finite radius.

For the two solutions that are outgoing at infinity, let $U$ collect their upper two components and $V$ their lower two components. A nonzero $V$ at finite radius does not mean that incoming radiation is imposed at infinity. Where $U$ is invertible, $Z=VU^{-1}$ describes the same allowed linear combinations without tracking the overall amplitudes of the two columns. It obeys
\begin{equation}\label{eq:Riccati}
Z'=C+D_4Z-ZA-ZB_{12}Z.
\end{equation}
The chain rule gives $dZ/dt=e^{i\theta}Z'$. At a transition point we reconstruct the linear solutions and continue to $r=r_m$. Appendix \ref{app:polar_riccati} gives the reduction and transport details.

\subsection{Matching functions and continuation}

At $r=r_m$, the axial inner and outer vectors must be proportional. The polar bases $Y_{\rm in}$ and $Y_{\rm out}$ must share a nonzero vector in the same four-dimensional constrained system. In terms of the corresponding matching matrices,
\begin{align}
D_{\rm ax}(\omega)
 &=\det[\bm y_{\rm in},\bm y_{\rm out}]=0,\\
D_{\rm pol}(\omega)
 &=\det[Y_{\rm in},-Y_{\rm out}]=0 .
\end{align}
We solve $\Re D=\Im D=0$ by a local Newton-type iteration. For fundamental modes the root-iteration tolerance is $10^{-9}$, the finite-difference function-error parameter is $10^{-12}$, and the nominal limit is 45 evaluations. We require convergence and $|D|<10^{-8}$ for the normalized determinant, and record the smallest matching singular value.

The main grids use the seed $\omega_{\rm seed}=\Zh\,\omega_{\rm vac}$. The smaller-mass and sound-speed scans use continuation from nearby roots. We vary the series orders, starting radii, matching radius, contour and integration tolerances in the refinement checks. The largest fundamental checkpoint shifts are $9.85\times10^{-12}\MBH^{-1}$ (axial) and $4.95\times10^{-12}\MBH^{-1}$ (polar). The independent vacuum comparison uses the Schwarzschild gravitational boundary problem, without an acoustic channel. Appendix \ref{app:numerics} gives the transport settings and validation.

\section{Fundamental modes}
\label{sec:fundamental}

\begin{figure*}[tbp]
\centering
\includegraphics[width=\textwidth]{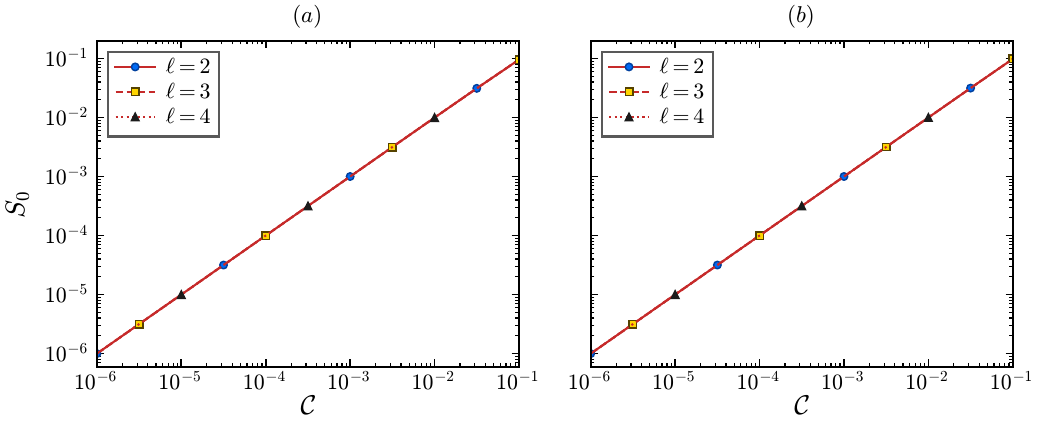}
\caption{Axial gravitational fundamental-mode shifts relative to the vacuum Schwarzschild frequency of the same multipole, as a function of $\Comp=M/a_0$. We hold $M=10\MBH$ (a) and $M=10^3\MBH$ (b) fixed while $a_0$ varies. Solid, dashed and dotted curves correspond to $\ell=2,3,4$. Lines connect the computed points on the Schwarzschild-connected axial gravitational branch.}
\label{fig:axial}
\end{figure*}

\begin{figure*}[tbp]
\centering
\includegraphics[width=\textwidth]{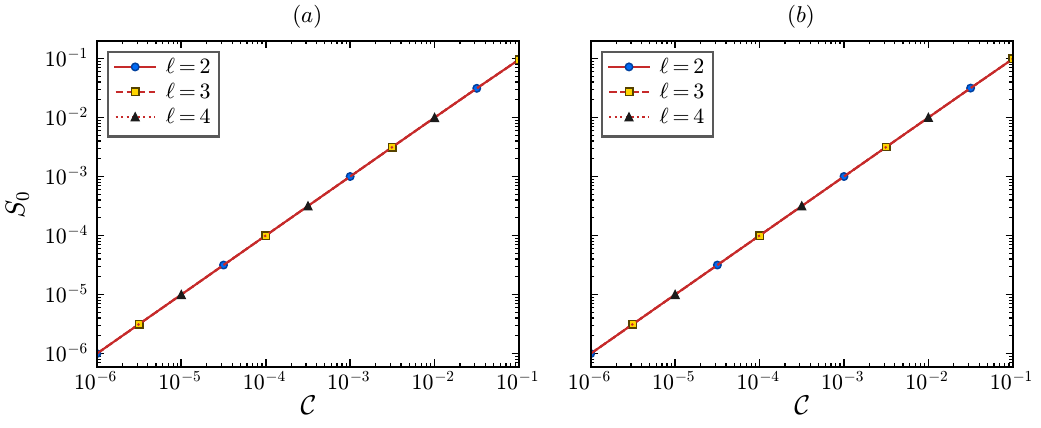}
\caption{Polar gravitational fundamental-mode shifts for the same masses, multipoles and compactness range as Fig.~\ref{fig:axial}. We impose $c_{s,r}=0.9$, $c_{s,t}=0$, retarded metric and acoustic conditions at the horizon, and outgoing metric and acoustic conditions at infinity, with no fluid cutoff. The similarity of the axial and polar shift magnitudes does not by itself establish isospectrality.}
\label{fig:polar}
\end{figure*}

Following the above methodology and numerical implementation we obtain the axial and polar QNMs $\omega_n$, where $n=0,1,2,\dots$ designates the overtone number of the mode. For example, $n=0$ is the fundamental mode that decays slower, and it is the one that dominates the ringdown signal at late-times before tail-effects take place \cite{Cardoso:2022whc}.

\subsection{Axial and polar redshift}

To benchmark the axial redshift, with respect to the results in Ref. \cite{Pezzella:2024tkf}, and calculate the polar QNMs and its redshift, we first define 
\begin{equation}
 R_n=\frac{\omega_{n}}{\omega_{{\rm vac},n}},\qquad
 S_n=\frac{|\omega_{n}-\omega_{{\rm vac},n}|}{|\omega_{{\rm vac},n}|}=|R_n-1|.
 \label{eq:redshift_ratio}
\end{equation}
We, also, define the absolute normalized difference between axial and polar modes
\begin{equation}
 \Delta_n=
 \frac{|\omega_{{\rm pol},n}-\omega_{{\rm ax},n}|}
      {|\omega_{{\rm ax},n}|},
 \label{eq:Deltan}
\end{equation}
in order to quantify a potential parity-isospectrality breaking. If the axial--polar isospectrality holds, then $\Delta_n=0$.

\begin{figure}[tbp]
\centering
\includegraphics[width=\columnwidth]{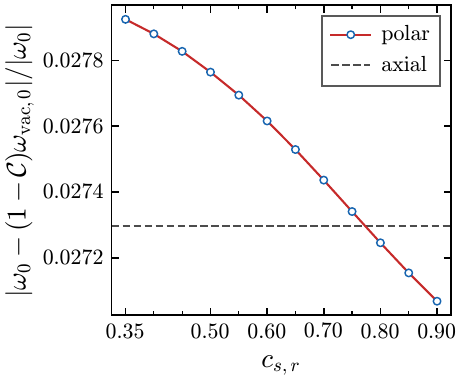}
\caption{Sound-speed dependence of the fundamental $\ell=2$ modes for $M=M_{\rm BH}$, $a_0=10M_{\rm BH}$, and $c_{s,t}=0$. The radial sound speed $c_{s,r}$ is sampled from $0.35$ to $0.90$ in steps of $0.05$. The fractional deviation from the first-order redshift prediction $(1-\mathcal{C})\omega_{\rm vac}$, normalized by the magnitude of the corresponding mode frequency is shown. The dashed axial result is independent of $c_{s,r}$. The largest sampled splitting is $7.38\times10^{-4}$ at $c_{s,r}=0.35$.}
\label{fig:soundspeed}
\end{figure}

We begin with the fundamental branches connected to the vacuum Schwarzschild $n=0$ QNM. Figures \ref{fig:axial} and \ref{fig:polar} show the axial and polar shifts for $M/\MBH=10,10^3$ and varying angular number $\ell=2,3,4$. At small $\Comp$, the plotted shift $S_0$ is proportional to the compactness and, at the scale of the figures, nearly independent of $\ell$. We fit $R_0=1-\sigma\Comp$ by unweighted complex least squares, with the intercept fixed at one. For the polar modes, the fits restricted to $\Comp\le10^{-3}$ give $\Re\sigma$ within $4.5\times10^{-4}$ of unity. Over the full range, the fitted slopes have small imaginary parts, so $1-\Comp$ is a good leading-order approximation rather than an exact real rescaling. We then measure the relative discrepancy from a redshift factor $Z=1-\Comp$ as $\delta_Z=|R_0-Z|/|R_0|$. Across the polar fundamental scans, the maximum discrepancy from $Z$ is $0.5\%$ for $M=10\MBH$ and $0.19\%$ for $M=10^3\MBH$. At $\Comp=10^{-1}$ and $M=10\MBH$, using the horizon lapse $Z=\Zh$ reduces the discrepancy $\delta_Z$ to $0.13\%$ in both sectors. The remaining discrepancy depends on the lapse reference and cannot be fully attributed to polar--fluid coupling alone. Thus, to test isospectrality, we next compare the axial and polar frequencies directly.

\subsection{Sound-speed dependence and parity splitting}

\begin{figure}[tbp]
\centering
\includegraphics[width=\columnwidth]{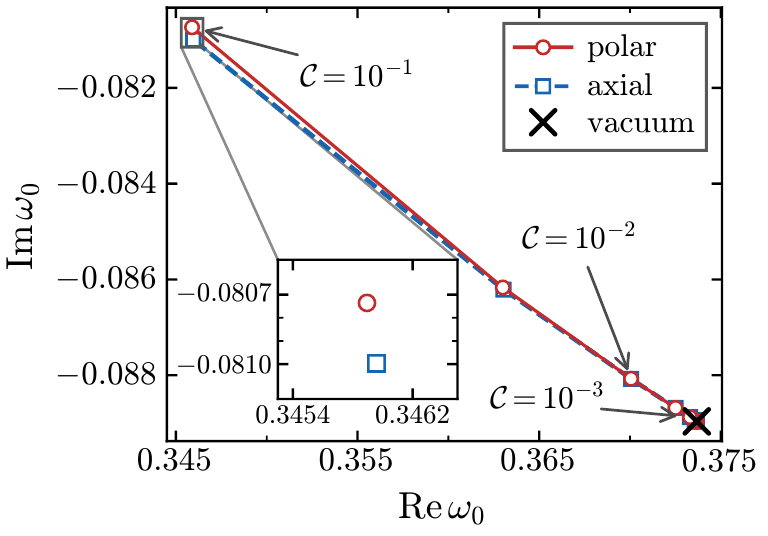}
\caption{Complex-frequency trajectories of the gravitational fundamental modes at $\MBH=M=1$, $\ell=2$, with polar $c_{s,r}=0.9$ and $c_{s,t}=0$. The compactness increases from $10^{-6}$ to $10^{-1}$. The black cross marks the vacuum frequency ($\Comp=0$), and labels identify selected compactness. The inset magnifies the $\Comp=10^{-1}$ endpoints. Frequencies are in units of $\MBH^{-1}$.}
\label{fig:traj0}
\end{figure}

\begin{figure*}[tbp]
\centering
\includegraphics[width=\textwidth]{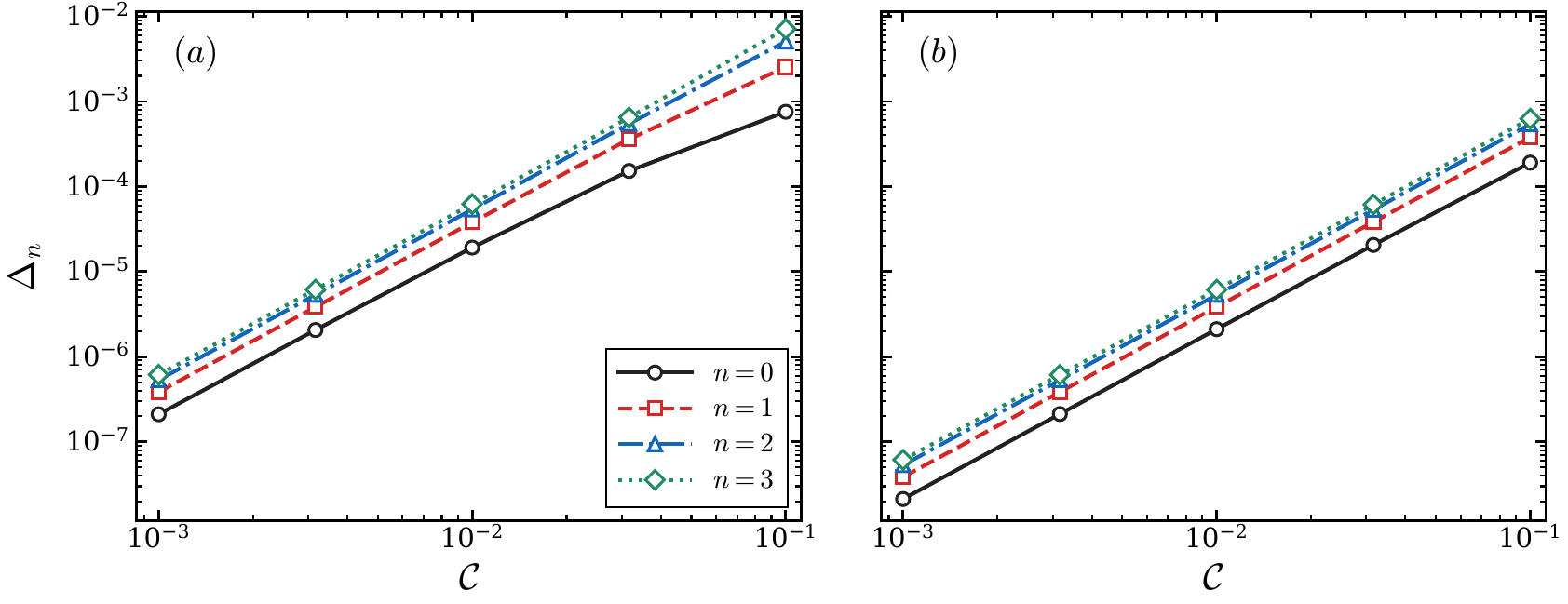}
\caption{Relative axial--polar splitting, Eq.~\eqref{eq:Deltan}, as a function of $\Comp=M/a_0$ for $\ell=2$, $c_{s,r}=0.9$ and $c_{s,t}=0$. Panels (a) and (b) have $M=\MBH$ and $M=10\MBH$. Each sequence holds $M$ fixed and varies $a_0$. Curves distinguish $n=0,1,2,3$. Lines connect computed samples and are not power-law fits. Both axes are logarithmic.}
\label{fig:splitn}
\end{figure*}

\begin{table*}[tb]
\caption{Axial and polar frequencies and parity gaps at $\Comp=10^{-1}$, $\ell=2$, and $c_{s,r}=0.9$. Here, $a_0/\MBH=10$ for $M/\MBH=1$ and $a_0/\MBH=10^2$ for $M/\MBH=10$. Frequencies are rounded for
display. The gaps use the unrounded complex frequencies.}
\label{tab:endpoints}
\begin{ruledtabular}
\begin{tabular}{ccccc}
$M/\MBH$ & $n$ & $\MBH\,\omega_{{\rm ax},n}$ &
$\MBH\,\omega_{{\rm pol},n}$ & $\Delta_n$\\
\hline
1  & 0 & $0.345958641-0.080996630i$ & $0.345895408-0.080735825i$ & $7.553\times10^{-4}$\\
1  & 1 & $0.320890-0.249249i$ & $0.321198-0.248263i$ & $2.540\times10^{-3}$\\
1  & 2 & $0.277730-0.435327i$ & $0.278524-0.432840i$ & $5.056\times10^{-3}$\\
1  & 3 & $0.230281-0.643059i$ & $0.230749-0.638216i$ & $7.122\times10^{-3}$\\
10 & 0 & $0.338084002-0.080249999i$ & $0.338033436-0.080207191i$ & $1.907\times10^{-4}$\\
10 & 1 & $0.313590-0.247123i$ & $0.313534-0.246983i$ & $3.782\times10^{-4}$\\
10 & 2 & $0.272081-0.431646i$ & $0.271990-0.431387i$ & $5.385\times10^{-4}$\\
10 & 3 & $0.227057-0.636670i$ & $0.226885-0.636283i$ & $6.269\times10^{-4}$\\
\end{tabular}
\end{ruledtabular}
\end{table*}

At $M=\MBH$, $a_0=10\MBH$ and $\ell=2$, the horizon lapse is $\Zh=0.91784$. Using it as a redshift factor in place of $1-\Comp$ reduces the discrepancy $\delta_Z$ from $2.7\%$ to $0.9\%$ in both sectors. Figure~\ref{fig:soundspeed} compares the redshift discrepancy with the parity splitting. Lowering $c_{s,r}$ from $0.9$ to $0.35$ shifts the polar frequency by $7.38\times10^{-4}|\omega_{\rm vac}|$, while leaving the axial reference unchanged. The lower radial sound speed limit is chosen so that the near-horizon acoustic branch point requirement $\Re\omega_0>1.05\,\omega_{\rm th}$ is met. The aforementioned requirement that we have set in our model is violated at $M_{\rm BH}\Re\omega_0\simeq0.346$, which corresponds to $c_{s,r}\simeq0.314$. This is the underlying reason that we begin our mode search at $c_{s,r}=0.35$. Therefore, at fixed halo parameters, changing $c_{s,r}$ leaves $Z_h$ and the axial frequency unchanged but moves the polar frequency. That change is attributable to the acoustic response in our model, including its boundary conditions. It shows that the fluid affects the polar mode beyond the common redshift, while leaving open how much of the splitting at any one sound speed comes from that response.

Finally, Figure~\ref{fig:traj0} follows both sectors in the complex plane. Increasing compactness lowers both $\Re\omega_0$ and $|\Im\omega_0|$. The magnified endpoint reveals a separation far smaller than the common redshift. The corresponding QNMs are shown in Table~\ref{tab:endpoints}. 

\section{Overtones}
\label{sec:overtones}

Here, we extend the axial-polar comparison to the first three QNM overtones, $n=1,2,3$. The labels refer to candidate branches associated with the corresponding vacuum Schwarzschild gravitational modes. We focus on compactness that spans $\Comp\in[10^{-3},10^{-1}]$, due to the fact that when $\Comp\leq 10^{-3}$, the effect of the non-vacuum QNMs is mostly a redshift. Here, we want to test the limits of parity-isospectrality splitting, which, as we will demonstrate, is enlarged for higher overtones and compact halo parameters.

\begin{figure*}[tbp]
\centering
\includegraphics[width=\textwidth,height=0.9\textheight,keepaspectratio]{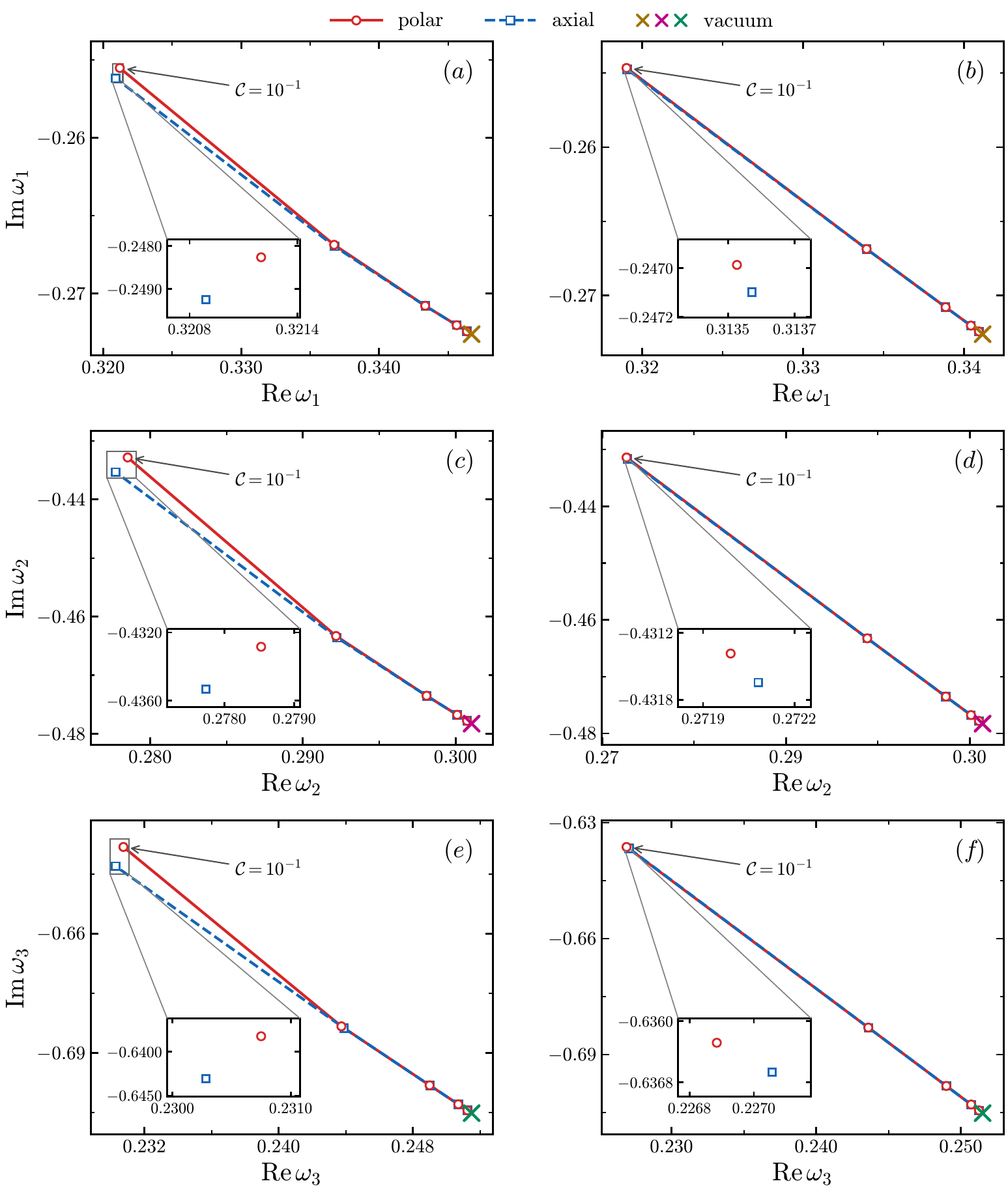}
\caption{QNM trajectories for the first three gravitational overtones at $\MBH=1$, $\ell=2$, $c_{s,r}=0.9$ and $c_{s,t}=0$. The left column (a, c, e) has $M=\MBH$ and the right column $M=10\MBH$ (b, d, f). Rows from top to bottom show $n=1,2,3$. Polar modes are shown with red circles joined by solid lines, while axial modes are shown with blue squares joined by dashed lines. Dark yellow, magenta, and emerald green crosses corresponds to the vacuum Schwarzschild $n=1,2$ and $3$ QNM frequencies. Moving away from the vacuum marker, the sampled compactness values are $10^{-3}$, $10^{-5/2}$, $10^{-2}$, $10^{-3/2}$ and $10^{-1}$, with $M$ fixed along each trajectory. Insets magnify the two $\Comp=10^{-1}$ endpoint modes. Frequencies are in units of $\MBH^{-1}$.}
\label{fig:trajn}
\end{figure*}

\subsection{QNM overtones and isospectrality}

We use $\ell=2$, $M/\MBH=1,10$, and the same closure as in Equation~\eqref{eq:closure}, with $c_{s,r}=0.9$ and $c_{s,t}=0$. Each sequence holds $M$ fixed and varies $a_0=M/\Comp$. The Schwarzschild reference for each overtone has been obtained from Ref. \cite{BertiRingdownData}, converted to the convention $e^{-i\omega t}$ and frequencies in units of $\MBH^{-1}$. Figure \ref{fig:splitn} shows $\Delta_n$ with respect to the compactness for the fundamental QNMs and first three overtones. We find that, at every sampled compactness, $\Delta_n$ increases with $n$, from the fundamental mode through the $n=3$ overtone. While this is true for both halo masses, we note that the lighter the halo, the larger the splitting. At $\Comp=10^{-1}$ the third-overtone gaps are $\sim10$ and $\sim 3.5$ times the fundamental-mode gaps for $M/\MBH=1$ and 10, respectively (see Table \ref{tab:endpoints}). At lower compactness the curves are close to quadratic in $\Comp$. For example, at $\Comp=10^{-3}$, $\Delta_n/\Comp^2$, with $n=1,2,3$, is about $(0.385,0.533,0.613)$ for $M=\MBH$ and ten times smaller for $M=10\MBH$. Since changing $M$ at fixed $\Comp$ also changes $a_0$, these samples do not establish a universal inverse-mass law.

\subsection{Overtone trajectories}

Figure~\ref{fig:trajn} follows the overtone branches in the complex plane. As compactness increases, both sectors oscillate more slowly and damp less rapidly. The polar modes are less strongly damped than their axial partners throughout the sampled grid. At $\Comp=10^{-1}$ the polar real parts are larger for $M=\MBH$ and smaller for $M=10\MBH$ at each of $n=1,2,3$. For the third overtone in the equal-mass case $M=\MBH$, damping accounts for about $99\%$ of the \emph{squared frequency separation}: the two frequencies are separated almost entirely in the imaginary direction. A larger splitting does not establish observability: the eigenfrequencies do not determine excitation amplitudes, and higher overtones damp more rapidly. For numerical resolutions tests of the overtones see Appendix \ref{app:overtone-settings} and \ref{app:followup_endpoints}.

\section{Discussion}
\label{sec:discussion}

\paragraph{Redshift and parity sensitivity.} A departure from a redshift law and a parity splitting are separate effects. A resolved nonzero $\Delta_n$ rules out a common factor
as an exact description of both sectors, even if it approximates each one well. The trend of the error with overtone number depends
on the redshift reference. For $M=\MBH$ and $\Comp=10^{-1}$, the error relative to $\Zh$ decreases from $0.86\%$ to $0.61\%$ from $n=0$
to $n=3$ in the axial sector, while it increases from $0.86\%$ to $1.33\%$ in the polar sector. To test a common rescaling directly, we compare overtone-to-fundamental ratios, which cancel any common factor. The third-overtone ratio changes by $1.39\%$ axially and $1.97\%$ in the polar sector relative to vacuum (see Appendix \ref{app:rescaling_audit}).

\paragraph{Role of the fluid.} The total axial--polar difference includes every difference between the perturbation systems and cannot be assigned to a single coupling term. In the model without a horizon cutoff, changing the axial prescription alone can shift the fundamental mode by more than the baseline parity gap. At $\Comp=10^{-1}$, the equal-mass fundamental splitting is larger at $c_s=0.95$ than at $0.9$ (see Appendix \ref{app:closure_audit}). The gravitational overtones studied here are distinct from the long-lived signal found in Ref. \cite{Cardoso:2022whc}, which used a varying sound speed and an inner fluid cutoff. Our auxiliary search did not establish a coupled fluid-mode pole. The compact examples are controlled tests of the model rather than estimates for typical galactic dark matter halos.

\section{Conclusions}\label{sec:conclusions}

We have calculated and compared axial and polar gravitational QNMs of a BH in a relativistic Hernquist-type halo, from the fundamental mode to the first three overtones connected to the vacuum Schwarzschild spectrum. The uncut polar calculation keeps both metric and acoustic branches at the horizon and at infinity, with constant radial sound speed, vanishing tangential sound speed and no inner matter cutoff. The vacuum benchmarks recover isospectrality within numerical resolution.

At small compactness, both sectors follow the redshift law $1-\Comp$. At larger compactness, the exact horizon lapse $\Zh=e^{\Upsilon(r_h)/2}$ accounts for most of the deviation from the redshift at the quoted precision, and the remaining discrepancy relative to this reference is comparable in the two parities at the fundamental mode. For $\ell=2$, $c_{s,r}=0.9$, $M/\MBH=1,10$ and $10^{-3}\le\Comp\le10^{-1}$, the parity splitting grows from $n=0$ through $n=3$ at every sampled compactness of the uncut model. At $\Comp=10^{-1}$ the third-overtone splittings are about $7.12\times10^{-3}$ and $6.27\times10^{-4}$ for the two mass families, larger than the fundamental ones by factors of $\sim 10$ and $\sim 3.5$. Overtone-to-fundamental frequency ratios depart from a common rescaling in both sectors. Within this model, the overtone spectrum, thus, carries a parity-dependent imprint of the halo that no common frequency rescaling can reproduce.

These conclusions rest on the polar--fluid closure and the acoustic boundary conditions. Retaining the splitting without the divergent acoustic branch, and testing it in a dynamically stable matter model remains necessary. Small collisionless Einstein--Vlasov shells around a Schwarzschild BH, for which linear stability has been established under spherically symmetric perturbations, could provide a starting point for testing the nonspherical response \cite{Jabiri:2020oke,Gunther:2022phn}. It would also be useful to extend the calculation to rotating remnants and follow the modes that reduce to the vacuum Kerr spectrum as the matter is removed \cite{Fernandes:2025osu}.

A calculation of mode excitation is needed to connect these spectral features to observable ringdown waveforms. The residues of the coupled Green function, together with a model for the perturbation produced by the merger, would give the amplitudes of the shifted gravitational modes and test whether a fluid-led contribution is detectable \cite{Berti:2006wq,Cardoso:2022whc}. This would establish whether the larger overtone splittings found here affect the part of
the signal that can be measured, given the faster damping of the overtones \cite{Berti:2022xfj,Spieksma:2024voy}.

For tests of vacuum general relativity, a common frequency rescaling can be absorbed into the inferred Kerr mass at fixed dimensionless spin. Ratios of several measured frequencies and damping rates, together with independent inspiral information where available, could help separate this common shift from mode-dependent effects \cite{LIGOScientific:2025wao}. Comparing vacuum Kerr templates with waveforms that include matter within general relativity, while allowing for uncertainty in the remnant mass and spin, would quantify when an environment can bias a vacuum test and when it can itself be constrained.

\begin{acknowledgments}
The authors would like to thank Prof. Andrea Maselli and Prof. Vitor Cardoso for valuable comments and fruitful discussions throughout the preparation of the manuscript.
K.D. acknowledges financial support provided by FCT--Funda\c{c}\~ao para a Ci\^encia e a Tecnologia, I.P., under the Scientific Employment Stimulus -- Individual Call -- Grant No.~2023.07417.CEECIND/CP2830/CT0008. K.D. would also like to thank the Funda\c{c}\~ao para a Ci\^encia e Tecnologia (FCT), Portugal, for the financial support to the Center for Astrophysics and Gravitation (CENTRA/IST/ULisboa) through grant No.~UID/PRR/00099/2025 and grant No.~UID/00099/2025.
This project has received funding from the European Union's Horizon MSCA-2022 research and innovation programme ``Einstein Waves'' under grant agreement No.~101131233.
\end{acknowledgments}

\section*{Data availability}
The figure data and source code are available upon reasonable request.

\appendix

\section{Polar equations and numerical method}\label{app:matrix}

\subsection{Fluid variables and the angular condition}
\label{app:physical_audit}\label{app:common_tilt}

In Regge--Wheeler gauge, $Y=Y_{\ell m}e^{\lambda t}$ and $\Omega_{AB}$ is the unit-sphere metric. The perturbations are $h_{tt}=aH_0Y$, $h_{tr}=H_1Y$, $h_{rr}=H_2Y/b$, and $h_{AB}=r^2K\Omega_{AB}Y$. The trace-free angular Einstein equation gives $H=H_0=H_2$ for $\ell\ge2$. For $\rho>0$, define the radial pressure, radial flux, and angular fluid amplitudes by
\begin{align}
 P&=s\delta\rho,&
 R&=\frac{aW}{\pi b(3b+1)},&
 J_{\rm ang}&=\frac{2\pi\rho H}{\lambda}.
\end{align}
The value of $J_{\rm ang}$ follows from angular momentum conservation. The velocity perturbations are
\begin{align}
 \delta u^t&=\frac{H}{2\sqrt a}Y,\qquad
 \delta u^r=\frac{bR}{\sqrt a\,\rho}Y,\nonumber\\
 \delta u^A&=\frac{\sqrt a J_{\rm ang}}{4\pi r^2(\rho+p_t)}\Omega^{AB}D_BY ,
\end{align}
where $D_A$ is the derivative on the unit sphere. We set $\delta k^A=0$. Normalization and orthogonality give $\delta k^r=-\sqrt b\,HY/2$ and $\delta k^t=\sqrt b\,H_1Y/a+\delta u^r/\sqrt{ab}$. The stress perturbations are
\begin{align}
 \delta T_{tt}&=a(\delta\rho-\rho H)Y,&
 \delta T_{tr}&=-(\rho H_1+R)Y,\nonumber\\
 \delta T_{rr}&=PY/b,&
 \delta T_{tA}&=-aJ_{\rm ang}D_AY/(4\pi),\nonumber\\
 \delta T_{rA}&=0,&\delta T_{AB}&=p_t h_{AB}.
\end{align}
The polar gauge has $h_{rA}=0$, so $\delta k^A=0$ is equivalent to $\delta k_A=0$. We impose $\delta k_A=0$ in the axial sector as well. There $\delta T^r{}_A=-p_t\sqrt b\,\delta k_A$ and $\delta T^t{}_A=(\rho+p_t)\delta u_A/\sqrt a$. Angular momentum conservation gives
\begin{align}
 \lambda\delta T^t{}_A&=-(\partial_r+d)\delta T^r{}_A,\nonumber\\
 d&=\frac{a'}{2a}-\frac{b'}{2b}+\frac{2}{r}.
\end{align}
For nonzero frequency and $\rho+p_t\ne0$, $\delta k_A=0$, therefore, implies $\delta u_A=0$. This gives $\delta T_{tA}=p_t h_{tA}$ and $\delta T_{rA}=p_t h_{rA}$, which is the covariant choice of Ref. \cite{Chakraborty:2024gcr}.

\subsection{Radial equations and constraint}\label{app:constraint}

For $\bm X=(H_1,H_0,K,W,\delta\rho)^T$, we use $D=-2m'$, $L=\ell(\ell+1)$, $s=c_s^2$, and $\lambda=-i\omega$. The nonzero entries of the metric rows of $F$ are
\begin{align}
 F_{11}&=-\frac{rb'-b+1}{2rb},&
 F_{12}&=\frac{\lambda^2r^2+aD}{\lambda r^2b},\nonumber\\
 F_{13}&=\lambda/b,&
 F_{21}&=\frac{La+2\lambda^2r^2}{2\lambda r^2a},\nonumber\\
 F_{22}&=\frac{2b-1}{rb},&
 F_{23}&=\frac{1-3b}{2rb},\nonumber\\
 F_{24}&=\frac{8a}{\lambda(3b+1)b},&
 F_{31}&=\frac{L}{2\lambda r^2},\nonumber\\
 F_{32}&=1/r,& F_{33}&=F_{23},&
 F_{34}&=F_{24}.
\end{align}
The nonzero entries of the two fluid rows are
\begin{align}
 F_{42}&=\frac{(-La+\lambda^2r^2)(3b+1)D}{16\lambda r^4 a},
 \nonumber\\
 F_{43}&=\frac{\lambda(3b+1)^2D}{32r^2ab},\nonumber\\
 F_{44}&=\frac{(9b+1)b'}{2b(3b+1)}-\frac{b+3}{2rb},\nonumber\\
 F_{45}&=-\frac{\pi\lambda(3b+1)}{a},\nonumber\\
 F_{51}&=\frac{D[-La(b+1)+4\lambda^2 r^2b]}
                 {64\pi\lambda r^4abs},\nonumber\\
 F_{52}&=-\frac{(3b-1)D}{32\pi r^3bs},\nonumber\\
 F_{53}&=\frac{(b+1)(3b-1)D}{64\pi r^3 b^2s},\nonumber\\
 F_{54}&=-\frac{4\lambda^2r^2b+a(b+1)D}
                  {4\pi\lambda r^2(3b+1)b^2s},\nonumber\\
 F_{55}&=-\frac{b(3s-1)+s+1}{2rbs}.
\end{align}
The constraint row $\bm c$ satisfies
\begin{equation}
 \bm c'+\bm c F=\frac{a'}{2a}\bm c .
 \label{eq:constraint_flow}
\end{equation}
Thus $\bm c\bm X\propto\sqrt a$ and stays zero if it vanishes at one radius. If $c_{H_0}\ne0$, Eq.~\eqref{eq:constraint_flow} also ensures that eliminating $H_0$ preserves its radial equation.

\subsection{Horizon data and integration from infinity}
\label{app:horizon}\label{app:polar_riccati}

Let us set $h=r_h$ and $A_h=a'(h)$. For the ingoing gravitational vector in the rescaled variables, we use $v_{g,H_0}=1$, $v_{g,H_1}=h\lambda/p_g$, and
\begin{equation}
 v_{g,K}=\frac{L v_{g,H_1}/(2\lambda h^2)+1/h}{p_g-1/2}.
\end{equation}
The fluid components follow from $(p_g I-R_0)\bm v_g=0$. The acoustic vector is $\bm v_f=(0,0,0,-\pi\lambda/[A_h(p_f+1)],1)^T$. The constraint excludes the additional exponent $p=1/2$ of the full system.

For the outer integration, we retain $\bm z=\bm X_{\mathcal J}$ with $\mathcal J=(H_1,H_0,W,\delta\rho)$. Substituting $K=-\bm c_{\mathcal J}\bm z/c_K$ gives $F^{(K)}=F_{\mathcal J\mathcal J}-F_{\mathcal JK}\bm c_{\mathcal J}/c_K$. The subscripts select rows and columns. For $\bm u=T_\infty\bm z$ with $T_\infty=\mathrm{diag}(r^{-1},r^{-1},r^2,r^2)$, the radial matrix is
\begin{equation}
 G_\infty=T_\infty F^{(K)}T_\infty^{-1}
            +T_\infty' T_\infty^{-1}.
\end{equation}
The wave basis lists outgoing gravitational and acoustic vectors first, then their incoming counterparts:
\begin{equation}
 S_\infty=
 \begin{pmatrix}
 \lambda&0&-\lambda&0\\
 -\lambda&0&-\lambda&0\\
 0&4\pi c_s&0&-4\pi c_s\\
 0&1&0&1
 \end{pmatrix}.
\end{equation}
The radial matrix in this basis is $\mathcal M=S_\infty^{-1}G_\infty S_\infty$. We write the two outgoing solutions as the columns of $\binom{U}{V}$. The lower block $V$ may be nonzero at finite radius. If $U$ is invertible, the columns of $\binom{I_2}{Z}$ with $Z=VU^{-1}$ give the same linear combinations. Differentiating $Z$ gives the Riccati equation in the main text (see Eq.~\eqref{eq:Riccati}). If $U$ is singular, the two solutions can instead be integrated directly.

Inward integration along $r=r_m+e^{i\theta}t$ uses decreasing $t$ with $dZ/dt=e^{i\theta}Z'$. Using $L_{\rm out}-t$ reverses the sign of the derivative. To switch to direct integration, we multiply $\binom{I_2}{Z}$ by $S_\infty$, undo $T_\infty$, and recover $K$ before continuing to $r_m$.

\section{Numerical settings and checks}\label{app:numerics}

\subsection{Integration settings and vacuum checks}\label{app:overtone-settings}

For the settings in Table~\ref{tab:prd_settings}, we set the outer contour length to the largest of the halo factor times $a_0$, the wave factor divided by $|\omega|$, and twice the transition distance. In the polar calculation, the switch distance is the contour distance from $r_m$ at which we tighten the tolerance as we integrate inward. Absolute tolerances are $0.01$ times the relative tolerances. The $\Comp=10^{-1}$ overtone endpoints use direct integration.

\begin{table*}
\caption{Baseline and refinement settings for the fundamental modes and for overtones at fixed $M$ and $\Comp<10^{-1}$. Lengths are given in units with $\MBH=1$. The overtone settings are for $n=1,2,3$, $\ell=2$, and $c_s=0.9$.}
\label{tab:prd_settings}
\begin{ruledtabular}
\begin{tabular}{lcccc}
 & \multicolumn{2}{c}{Fundamental modes} & \multicolumn{2}{c}{Overtones} \\
Quantity & Baseline & Refinement & Baseline & Refinement \\
\hline
Horizon order $N_h$ & 6 & 8 & $12+2n$ & $14+2n$ \\
Horizon offset $x/r_h$ & $10^{-4}$ & $3\times10^{-4}$ & 0.04 & 0.05 \\
Outer order $N_\infty$ & 10 & 12 & 10 & 12 \\
Matching radius $r_m$ & 8 & 10 & 5 & 5.5 \\
Contour angle $\theta$ & 0.50 & 0.65 & 1.45 & 1.50 \\
Halo and wave factors & $(6,50)$ & $(9,70)$ & $(6,60)$ & $(9,80)$ \\
Transition distance & 60 & 100 & 45 & 65 \\
Relative tolerance (linear) & $2\times10^{-11}$ & $2\times10^{-13}$ & $7\times10^{-13}$ & $3\times10^{-13}$ \\
Relative tolerance (Riccati, closer to $r_m$) & $10^{-10}$ & $10^{-12}$ & $10^{-11}$ & $2\times10^{-12}$ \\
Polar tolerance switch distance & $10^4$ & $2\times10^4$ & $2\times10^4$ & $2\times10^4$ \\
Relative tolerance (far polar) & $10^{-7}$ & $2\times10^{-8}$ & $2\times10^{-8}$ & $2\times10^{-8}$
\end{tabular}
\end{ruledtabular}
\end{table*}

For overtones, we divide $D$ by its magnitude at the initial guess, with a minimum divisor of $10^{-30}$. The root tolerance is $2\times10^{-9}$, the finite-difference error parameter is $10^{-10}$, and the nominal limit is 55 evaluations. Only the initial guess for refinement is shifted by $10^{-4}(1+0.7i)\MBH^{-1}$ from the baseline root.

We estimate the frequency correction as $e_N=\|\mathbb J^{-1}(\Re D,\Im D)^T\|_2$. The Jacobian $\mathbb J$ uses centered steps of $10^{-5}\MBH^{-1}$ in $\Re\omega$ and $\Im\omega$. The error scale is the largest of $e_N$, the change under refinement, the differences from direct integration and the vacuum benchmark, and a minimum of $3\times10^{-12}\MBH^{-1}$. The direct-integration check raises this minimum to $1.19\times10^{-9}\MBH^{-1}$ for the new $M=\MBH$, $n=3$ samples.

In vacuum, the polar and Zerilli calculations agree with the axial fundamentals for $\ell=2,3,4$ to a few $10^{-12}\MBH^{-1}$. The largest axial and polar differences from the vacuum benchmarks are $(5.14,3.57)\times10^{-12}\MBH^{-1}$ for $n=1$ and $(2.17,2.15)\times10^{-10}\MBH^{-1}$ for $n=2$. For $n=3$, moving the match inward gives differences of $(1.41,2.97,3.05)\times10^{-12}\MBH^{-1}$ in the axial, Zerilli, and polar calculations. There is no acoustic condition at $M=0$. The limit $\Comp\to0$ at fixed $M$ instead leaves $M_{\rm ADM}$ unchanged.

\subsection{Equation and frequency checks at nonzero halo mass}
\label{app:followup_checks}\label{app:followup_endpoints}

We check the linearized Einstein and stress-energy conservation equations using Appendix~\ref{app:physical_audit}. We use Eqs.~(69)--(77) of the preprint of Ref. \cite{Chakraborty:2024gcr} with $p_r=0$ and our pressure law. For example, radial conservation requires
\begin{align}
0={}&P'+\frac{\lambda R}{a}
 +\left(\frac{2}{r}+\frac{g}{2}\right)P+\frac{g}{2}\delta\rho
 \nonumber\\
 &+\frac{\lambda\rho H_1}{a}
 -\frac{\rho H'}{2}-\frac{\rho rgK'}{4},
\end{align}
where $g=a'/a$.

The columns of $N$ satisfy $\bm cN=0$ and span the allowed field values. We use $FN$ and $(F'+F^2)N$ for the first and second derivatives. Each residual norm is divided by the sum of the norms of its terms. All nine checks give values below $10^{-11}$ at twelve radii in $2.05\le r/\MBH\le800$ for each polar endpoint. The matched fields differ by less than $2.39\times10^{-14}$ in relative norm. Their numerical derivatives give residuals up to $2.09\times10^{-8}$.

Quadrupling the outer length reduces the largest acoustic series residual from $4.29\times10^{-3}$ to $3.75\times10^{-9}$ at $c_s=0.9$ and from $9.15\times10^{-2}$ to $3.64\times10^{-6}$ at $c_s=0.95$. For the eight $c_s=0.9$ modes, the smallest singular values at the longer length are at most $6.20\times10^{-13}$. These residuals do not measure the error in the frequencies.

Using the largest spread found in the earlier endpoint runs, $e_\star=1.81\times10^{-6}\MBH^{-1}$, we count a frequency difference as resolved if
\begin{equation}
 |\omega_{{\rm pol},n}-\omega_{{\rm ax},n}|>20 e_\star .
 \label{eq:endpoint_screen}
\end{equation}
All eight pairs at $\Comp=10^{-1}$ pass by a factor of at least 1.83. Two direct integrations agree within $2.08\times10^{-9}\MBH^{-1}$. The $M=10\MBH$, $n=1$, $\Comp=10^{-3}$ gap fails the test. The rest of the earlier grid cannot be reassessed without its run records. For $M=\MBH$, we find $\Delta_1<\Delta_2<\Delta_3$ at ten values from $\Comp=10^{-1}$ to $10^{-1.5}$. This does not rule out close approaches between sampled points or nearby fluid modes.

\subsection{Horizon behavior and changes in the matter model}
\label{app:acoustic_audit}\label{app:closure_audit}

In ingoing time $v=t+r_*$, the acoustic density term is proportional to $e^{-i\omega v}x^{p_f+i\omega\tau_h}$. The real part of the radial exponent lies between $-1.22$ and $-1.08$, while the equilibrium density vanishes as $x$. The matched solutions have nonzero acoustic amplitudes, so their density perturbation diverges at the horizon. Dropping that solution would change the boundary problem.

At $\Comp=10^{-1}$, switching from the down to the up axial potential \cite{Chakraborty:2024gcr} changes the fundamental by $2.59\times10^{-3}|\omega_{\rm ax}|$ for $M=\MBH$ and by $4.67\times10^{-4}|\omega_{\rm ax}|$ for $M=10\MBH$. Both shifts exceed the baseline axial--polar gaps. We keep the polar equations fixed in this check.

At $c_s=0.70$, $0.90$, and $0.95$, the endpoint gaps increase with overtone number. The $M=10\MBH$, $c_s=0.70$, $n=0$ gap does not pass Eq.~\eqref{eq:endpoint_screen}. Refining the sound-speed runs changes the polar roots by at most $7.11\times10^{-7}\MBH^{-1}$. For $M=\MBH$ the fundamental gap at $c_s=0.95$ is $8.59\times10^{-4}$, larger than at $0.90$.

\subsection{Redshift and overtone ratios}\label{app:rescaling_audit}

Let $X=\{\mathrm{axial},\mathrm{polar}\}$ denote the axial or polar sector, and let $n=0$ denote the fundamental mode and $n\geq1$ the overtones. At fixed halo parameters and multipole $\ell$, we define
\begin{equation}
R_{X,n}=\frac{\omega_{X,n}}{\omega_{{\rm vac},n}},
\end{equation}
where $\omega_{{\rm vac},n}$ is the Schwarzschild frequency with the same $\ell$ and $n$. An exact rescaling by a positive real redshift factor would make $R_{X,n}$ real and positive. Instead, at $M=\MBH$ and $\Comp=10^{-1}$, the axial fundamental has $\arg R_{\mathrm{axial},0}=3.75\times10^{-3}\,\mathrm{rad}$. With $Z(r)=e^{\Upsilon(r)/2}$ and $\Zh=Z(r_h)$, replacing $\Zh$ by $Z(3.28\MBH)$ reduces the relative discrepancy $\delta_Z^\mathrm{axial}$ from $0.00861$ to $0.00376$. Here the discrepancy is $\delta_Z^\mathrm{axial}=|R_{\mathrm{axial},0}-Z_{\rm ref}|/|R_{\mathrm{axial},0}|$ for the chosen reference factor $Z_{\rm ref}$. The nonzero phase accounts for nearly all of the remaining discrepancy.

To test whether all modes in a given sector share the same frequency rescaling, we compare each overtone with the fundamental mode in that sector:
\begin{equation}
\begin{aligned}
I_{X,n}
  = \frac{R_{X,n}}{R_{X,0}}-1 
  = \frac{\omega_{X,n}/\omega_{X,0}}
           {\omega_{{\rm vac},n}/\omega_{{\rm vac},0}}-1 .
\end{aligned}
\label{eq:invariant_ratio}
\end{equation}
Any factor common to the overtone and the fundamental mode cancels, even if it is complex. Thus $100|I_{X,n}|$ is the percentage by which the complex overtone-to-fundamental frequency ratio differs from its Schwarzschild value. At $\Comp=10^{-1}$, the third-overtone ratio ($n=3$) differs from its Schwarzschild value by $1.39\%$ in the axial sector and $1.97\%$ in the polar sector for $M=\MBH$. For $M=10\MBH$, the corresponding differences are $0.205\%$ and $0.252\%$. In either sector, applying the rescaling inferred from the fundamental mode to the Schwarzschild third overtone does not recover the calculated halo frequency. This means that the factor that maps the vacuum fundamental mode to the halo fundamental mode does not also map the vacuum third overtone to the halo third overtone.

\bibliography{references}

\end{document}